\documentstyle[12pt]{article}

\begin{document}

\begin{center}
 {\Large \bf Cabibbo-suppressed non-leptonic B- and D-decays involving
tensor mesons}\\
\vspace{1.3cm}

{\large  J. H. Mu\~noz$^1$, A. A. Rojas$^1$ and G. L\'opez Castro$^2$}

$^1$ {\it Departamento de F\'{\i}sica, Universidad del Tolima,}\\
{\it A. A. 546, Ibagu\'e, Colombia} 

 $^2$ {\it Departamento de F\'{\i}sica, Centro de Investigaci\'on y de
Estudios}\\ {\it Avanzados del IPN, Apdo. Postal 14-740, 07000 M\'exico,
D.F., M\'exico}

\end{center}
\vspace{1.3cm}
\begin{abstract}
The Cabibbo-suppressed non-leptonic decays of $B$ (and $D$) mesons to
final states involving tensor mesons are computed using the
non-relativistic quark model of Isgur-Scora-Grinstein-Wise with the
factorization hypothesis. We find that some of these $B$ decay modes,
as $B \rightarrow (K^*, D^*)D^*_2$\ , can  have branching ratios as large
as  
$6 \times 10^{-5}$ which seems to be at the reach of future $B$ factories.
 \end{abstract}

\vspace{1.3cm}
PACS number(s): 13.15.Hw, 12.39.Jh, 14.40.Nd

\newpage

\begin{center}
\large \bf 1. Introduction
\end{center}

  $B$ meson factories at SLAC and KEK will soon start their operation.
Besides the central interest on the study of CP violation in the $B$
system, the precision of many properties of $B$ mesons that have already
been measured is expected to be improved there. Some suppressed decays of
$B$ mesons, either modes occurring at tree-level and 
suppressed by CKM factors or modes suppressed by dynamical effects, 
will certainly be accessible at these experiments for the first time. 
  Another kind of suppressed $B$ decays correspond to the modes containing
mesons that are radial or orbital excitations of the $q\bar{q'}$ system.
Semileptonic $B$ decays containing orbital excitations of the $c\bar{q}$
system as $D_1$ and $D_2^*$ mesons have been observed recently by CLEO
\cite{cleo21}, concluding that they can account for up to 20 \% of the $B$
semileptonic rate. The study of these decays is interesting to probe the
specific predictions for the hadronic matrix elements in the context of
 phenomenological quark models \cite{isgur} and heavy quark
effective theory \cite{wise,ebert,matsuda}.

   In a recent paper \cite{herman} we have computed the Cabibbo-favored
non-leptonic decay modes of $B$ mesons of the form $B \rightarrow PT,\,
VT$, where $P(V)$ is a pseudoscalar (vector) meson and the spin-2
tensor meson $T$ corresponds to the $p$-wave of the quark-antiquark
system.
We have found \cite{herman} that some of these decay modes have branching
ratios large enough to be observed in future measurements. Similar
conclusions have been reached in refs. \cite{kv_tensor,neubert_t}. 

In the present paper we consider the Cabibbo-suppressed two-body
non-leptonic decay modes of $B$ mesons that contain a light- or  
charmed-tensor meson in the 
final state. Despite additional Cabibbo-suppression factors, the
amplitudes for some of
these decays can be enhanced because they are favored by contributions
proportional to the $a_1$ QCD coefficient which appears in the effective
weak Hamiltonian and/or have more phase-space
available. We make use of the
non-relativistic quark model of ref. \cite{isgur} to evaluate the relevant
hadronic matrix elements of the $B \rightarrow T$ transitions, where $T$
is a light or heavy ({\it i.e.} charmed) tensor meson. 
For
completeness, we also compute the Cabibbo-suppressed two-body non-leptonic
$D$ decays involving tensor mesons that are allowed by phase space
considerations. The corresponding Cabibbo-favored $D$ decays have been
computed in ref. \cite{katoch1}. 

Let us mention that the 
$B \rightarrow T$ hadronic matrix element computed in ref. \cite{isgur}
has been used recently to evaluate the semileptonic rate of the
$B\rightarrow D_2^* l
\nu$ \cite{matsuda} decay mode. The heavy quark effective theory also
allows
a computation of the $B\rightarrow T$ matrix element and has been used to
evaluate the decay rates of the $B^- \rightarrow D_2^{*0} \pi^-$
\cite{neubert_t} (see also \cite{wise}) and $B\rightarrow D_2^* l \nu$
\cite{wise,ebert,matsuda} decays. 

Besides the interest of heavy meson
decays to tensor mesons in order to test properties of quark models or
symmetries of QCD for heavy quarks, one should mention that tensor mesons
({\it i.e.} $q\bar{q'}$ states with $L=1, \ S=1$ and $J^P=2^+$) belong to
one of the better established 16-plet under flavor SU(4).
 Indeed, according to the compilation of the Particle Data Group
\cite{pdg}, the following members of the 16-plet of tensor mesons have
already been observed: the isovector $a_2(1320)$ state, the isoscalars
$f_2(1270),\ f'_2(1525)$ and $\chi_{c2}(3556)$, the strange isospinor
$K_2^*(1430)$ and the charm isodoublet $D_2^*(2460)$ states. Although
there is not compelling evidence yet for the charmed-strange tensor meson, 
according to ref. \cite{pdg98} the $D_{sJ}^*(2573)$ has the width and
decay modes consistent with a $J^P=2^+$ $c\bar{s}$ state.

   Despite their low branching fractions, the observation of some
non-leptonic Cabibbo-suppressed $B$ and $D$ decays to lowest lying mesons
have been reported recently. For example, the following decay modes have
been observed: $B^- \rightarrow
D^0K^-$ \cite{cleo1}, $B^0 \rightarrow D^{*+}D^{*-}$ \cite{cleo2}, 
$D^0 \rightarrow K^-K^*,  \pi^+\pi^-$ \cite{E791} 
and $D^+ \rightarrow (\eta,\eta^{'})\pi^+, (\eta, \eta^{'})\rho^+$
\cite{cleo3}. As will be shown below, some Cabibbo-suppressed $B$ decays to
tensor mesons have branching ratios of order $10^{-5} \sim 10^{-6}$ which
look not too far from experimental searches at $B$ factories.

  The rest of the paper is organized as follows. In section 2 we write 
the effective non-leptonic weak Hamiltonians for Cabibbo-suppressed $B$
and $D$ decays and provide a classification for these decays. In section 3
we set our convention for mixing of octet and singlet states of SU(3) and
provide the numerical values of the parameters required for our
calculations. Our conclusions are given in section 4. Let us note that we
closely follow the notation and formulae obtained in ref. \cite{herman}

\

\begin{center}
\large \bf 2. Effective weak Hamiltonians for Cabibbo-suppressed $B$ and
$D$ decays
\end{center}

  The $B$ and $D$ decays of our interest are such that only one single
Cabibbo suppression factor occurs at a time.
The effective weak Hamiltonian for single Cabibbo-suppressed n on-leptonic
$B$ decays can be written as follows:
\begin{eqnarray}
{\cal H}_{eff}(\Delta b)&=& \frac{G_F}{\sqrt{2}}\left\{ V_{ub}V_{ud}^*
[a_1(\overline{u}b)(\overline{d}u)+a_2(\overline{d}b)(\overline{u}u)]
\right. \nonumber \\
&& \ \ \left. +
 V_{ub}V_{cs}^*[a_1(\overline{u}b)(\overline{s}c)+ 
a_2(\overline{s}b)(\overline{u}c)] \right. \nonumber \\
&& \ \  \left. +V_{cb}V_{cd}^*
 [a_1(\overline{c}b)(\overline{d}c)+a_2(\overline{d}b)(\overline{c}c)]
\right. \nonumber \\
 && \ \ \left. +
 V_{cb}V_{us}^*[a_1(\overline{c}b)(\overline{s}u)+
a_2(\overline{s}b)(\overline{c}u)] \right\} + h.c.  
\end{eqnarray}
where $(\overline{q}q^{'})$ is a short notation for the $V-A$ current,
$G_F$ denotes the Fermi constant, and $V_{ij}$ are the relevant CKM
mixing factors. In this paper we will take the following numerical values
for the QCD coefficients: $a_1=1.15,\ a_2=0.26$ \cite{cleo5}.

  In order to provide a classification for the wide set of these decays,
we will call {\it type I} decays those occurring through the first two
terms within curly brackets ({\it i.e.} proportional to $V_{ub}$) while
those proportional to $V_{cb}$ will be called of {\it type II}. Based on
current values of CKM matrix elements, one would naively expect that type
I $B$ decay branching ratios are suppressed by the factor
$|V_{ub}/V_{cb}V_{us}|^2 \approx 0.13$ with respect to type II decays.
Among
type I and II decays we will also distinguish between processes with
$\Delta s=0,\, 1$ associated to the change of strangeness in the second
weak vertex.

   In a similar way, the effective weak Hamiltonian for single
Cabibbo-suppressed $D$ decays is given by
\begin{equation}
{\cal H}_{eff}(\Delta c)=\frac{G_F}{\sqrt{2}}V_{cd}V_{ud}^*
\{a_1(\overline{d}c)(\overline{u}d)+a_2(\overline{u}c)(\overline{d}d)\}+h.c.
\end{equation}
where the numerical values for the QCD coefficients will be taken as
$a_1=1.26$ and $a_2=-0.51$ \cite{wirbel}.
Notice that we have included only the terms relevant for $D$ decays with
$\Delta s=0$ (type I), because type II transitions  are very suppressed
by phase space considerations. Note that $D \rightarrow VT$ are
completely forbidden by kinematics.

  Observe that due to the vector nature of the effective hadronic weak
currents in Eqs. (1,2), the matrix element $\langle T|\bar{q} q'|0
\rangle$
vanishes identically. Therefore, as already discussed in ref.
\cite{herman}, only one operator in the effective weak Hamiltonian will
contribute to the decay amplitude of a given process, {\it i.e.} the
amplitudes for our processes become proportional to either $a_1$ or $a_2$ 
alone.

\

\begin{center}
\large \bf 3. Mixing of states, decay constants and results
\end{center}

In this section we provide our convention for SU(3) octet-singlet mixing
of states and the numerical values for the decay constants required to
describe the $\langle P(V) |(\bar{q}q')|0 \rangle$ matrix elements.

  As is known, the breaking of SU(3) flavor symmetry produces a mixing
between octet and singlet states of SU(3) with $I=0$. In our convention,
this mixing leads to the following expressions for the physical states: 
\begin{eqnarray}
\eta &=& \frac{1}{\sqrt{2}}(u \overline{u} + d \overline{d}) 
\sin\phi_P - (s\overline{s})\cos\phi_P, \nonumber \\
\eta^{'} &=& \frac{1}{\sqrt{2}}(u \overline{u} + d \overline{d}) 
\cos\phi_P + (s\overline{s})\sin\phi_P, \nonumber \\
\omega &=& \frac{1}{\sqrt{2}}(u \overline{u} + d \overline{d}) 
\cos\phi_V + (s\overline{s})\sin\phi_V, \nonumber \\
\phi &=&  \frac{1}{\sqrt{2}}(u \overline{u} + d \overline{d}) 
\sin\phi_V - (s\overline{s})\cos\phi_V, \nonumber \\
f_2 &=&  \frac{1}{\sqrt{2}}(u \overline{u} + d \overline{d}) 
\cos\phi_T + (s\overline{s})\sin\phi_T, \nonumber \\
f_2^{'} &=& \frac{1}{\sqrt{2}}(u \overline{u} + d \overline{d}) 
\sin\phi_T - (s\overline{s})\cos\phi_T,
\end{eqnarray}
where the mixing angle is given by $\phi_i = \arctan{(1/\sqrt{2})} -
\theta_i$ ($i=P,V$ o $T$) and the experimental values of $\theta_i$ are
given by $-20^0,\ 39^0$ and $28^0$ \cite{pdg} for pseudoscalar ($\eta,
\eta'$), vector ($\omega,\phi$) and tensor ($f_2,f'_2$) mesons,
respectively.

   With the above convention, we have computed the decay amplitudes for
type I and II $B \rightarrow P(V)+T$ decays and the results are given in
the third column of Tables 1--4. The analogous results for type I
$D\rightarrow P+T$ transitions are given in the third column of Table 5.
As already mentioned,
these amplitudes are proportional to only one QCD coefficient appearing
in the weak Hamiltonians. The explicit expressions for the functions 
${\cal F}^{i \rightarrow f}$ and ${\cal F}^{i \rightarrow f}_{\mu\nu}$
appearing in the decay amplitudes 
and the properties of the symmetric polarizations tensors
$\epsilon_{\mu\nu}$ describing the spin 2 particles can be found in ref.
\cite{herman}.

   In order to provide numerical values of the branching ratios we use the
expressions for the decay rates given in Eqs. (9) and (11) of ref.
\cite{herman} and the following values of the CKM matrix elements 
\cite{pdg}:  $|V_{ub}|=3.3
\times 10^{-3}$, $|V_{ud}|=0.9740$, $|V_{cs}|=0.975$, $|V_{cb}|=0.0395$,
$|V_{cd}|=0.224$ and $|V_{us}|=0.2196$. The values for the lifetimes of
$B$ and $D$ mesons are taken from ref. \cite{pdg}. 

The decay constants of pseudoscalar mesons $f_P$ ( given in GeV units)
have the following central values:
$f_{\pi^-}=0.131$ \cite{pdg},
$f_{\pi^0}=0.130$
\cite{pdg}, $f_{\eta}=0.131$ \cite{neubert}, $f_{\eta^{'}}=0.118$
 \cite{neubert}, $f_{D_s}=0.280$ \cite{cleo4}, $f_D=0.
252$, $f_{\eta_c}=0.393$ \cite{sharma1} and $f_{K^+}=0.159$ \cite{pdg}.
$f_D$ is obtained using the theoretical prediction $f_D/f_{D_s}=0.90$
\cite{bernard} and the value for $f_{D_s}$. On the other hand, the central
values for the dimensionless decay constants of vector mesons $f_V$ 
are \cite{sharma1}: $f_{\rho}=0.281$, $f_{\omega}=0.249$,
$f_{\phi}=0.232$, $f_{D_s^{*}}=0.128$, $f_{D^{*}}=0.124$,
$f_{J/\psi}=0.1307$ and $f_{K^{*}}=0.248$.

   The branching ratios for Cabibbo-suppressed $B$ and $D$ decays
involving tensor mesons are given in the last column of Tables 1--5.

\

\begin{center}
\large \bf 4. Conclusions.
\end{center}

  Semileptonic $B$ decays to final states containing orbital excitations
of the
$q\bar{q'}$ system have already been observed in recent experimental
searches. These suppressed decay modes are expected to provide additional
tests of the QCD dynamics exhibited by phenomenological quark models or
the Heavy Quark Effective Theory predictions for the hadronic matrix
elements involving higher excitations of the $q\bar{q'}$ system.

  Based on the non-relativistic quark model of ref. \cite{isgur}, in this
paper we have computed the Cabibbo-suppressed decay modes of $B$ (and $D$)
mesons to final states involving $J^P=2^+$ tensor mesons. As observed in
Table 4, some of these $B$ decays as $B^- \rightarrow (D^{*-},\,
K^{*-}) D^{*0}_2$ and $\overline{B^0} \rightarrow (D^{*-},\,
K^{*-})D^{*+}_2$ can have branching ratios as large as $6 \times 10^{-5}$,
which seems to be at the reach of future $B$ factories. Despite the fact
that these modes have a Cabibbo-suppression factor, they exhibit branching
fractions comparable to some corresponding Cabibbo-allowed $B$ decays
(see for example, \cite{herman,kv_tensor}) 
because they are proportional to the $a_1$ coefficients (instead of $a_2$)
 and the available phase space is larger. Regarding $D\rightarrow PT$
transitions, the most favored  decay modes correspond to $D \rightarrow
\pi^+ (a_2, f_2)$ with branching fractions in the range $ (4 \sim 8)\times
10^{-6}$.

\begin{center}
\large Acknowledgements
\end{center}

  J. H. M. is grateful to Colciencias (Colombia) for financial support.

\newpage

\begin{center}
\begin{tabular}{|c|c|c|c|}
\hline
& Process & Amplitude $\times (V_{ub}V_{ud}^*)$ & $Br(B \rightarrow PT)$\\
\hline\hline
& $B^- \rightarrow \pi^- a_2^0$ & $a_1f_{\pi^-}{\cal F}^{B \rightarrow
a_2}(m^2_{\pi^-})/\sqrt{2}$ & $3.02\times 10^{-7}$ \\
& $B^- \rightarrow \pi^- f_2$ & $a_1f_{\pi^-}\cos\phi_T{\cal F}^{B
\rightarrow f_2}(m^2_{\pi^-})/\sqrt{2}$ & $3.25\times 10^{-7}$ \\
& $B^- \rightarrow \pi^- f_2^{'}$ & $a_1f_{\pi^-}\sin\phi_T{\cal F}^{B
\rightarrow f_2^{'}}(m^2_{\pi^-})/\sqrt{2}$ & $3.23\times 10^{-9}$ \\
& $B^- \rightarrow \pi^0 a_2^-$ & $a_2f_{\pi^0}{\cal F}^{B \rightarrow
a_2}(m^2_{\pi^0})/\sqrt{2}$ & $1.52 \times 10^{-8}$ \\
& $B^- \rightarrow \eta a_2^-$ & $a_2f_{\eta}\sin\phi_P{\cal F}^{B
\rightarrow a_2}(m^2_{\eta})/\sqrt{2}$ & $1.05 \times 10^{-8}$\\
& $B^- \rightarrow \eta^{'} a_2^-$ & $a_2f_{\eta^{'}}\cos\phi_P{\cal
F}^{B \rightarrow a_2}(m^2_{\eta^{'}})/\sqrt{2}$ & $4.18 \times 10^{-9}$
\\
& $\overline{B^0} \rightarrow \pi^- a_2^+$ & $a_1f_{\pi^-}{\cal F}^{B
 \rightarrow a_2}(m^2_{\pi^-})$ & $5.71 \times 10^{-7}$\\
$\Delta s =0$ & $\overline{B^0} \rightarrow \pi^0 a_2^0$ &
$-a_2f_{\pi^0}{\cal F}^{B \rightarrow a_2}(m^2_{\pi^0})/2$ & $7.18 \times
10^{-9}$ \\
& $\overline{B^0} \rightarrow \pi^0 f_2$ & $a_2f_{\pi^0}\cos\phi_T{\cal
F}^{B \rightarrow f_2}(m^2_{\pi^0})/2$ & $7.72 \times 10^{-9}$ \\
& $\overline{B^0} \rightarrow \pi^0f_2^{'}$ &
$a_2f_{\pi^0}\sin\phi_T{\cal F}^{B \rightarrow f_2^{'}}(m^2_{\pi^0})/2$ &
$7.68 \times 10^{-11}$ \\
& $\overline{B^0} \rightarrow \eta a_2^0$ & $-a_2f_{\eta}\sin\phi_P{\cal
F}^{B \rightarrow a_2}(m^2_{\eta})/2 $ & $4.98 \times 10^{-9}$ \\
& $\overline{B^0} \rightarrow \eta f_2$ &
$a_2f_{\eta}\sin\phi_P\cos\phi_T{\cal F}^{B \rightarrow
f_2}(m^2_{\eta})/2$  & $5.36 \times 10^{-9}$\\
& $\overline{B^0} \rightarrow \eta f_2^{'}$ &
$a_2f_{\eta}\sin\phi_P\sin\phi_T{\cal F}^{B \rightarrow
f_2^{'}}(m^2_{\eta})/2$ & $5.31 \times 10^{-11}$ \\
& $\overline{B^0} \rightarrow \eta^{'} a_2^0$ &
$-a_2f_{\eta^{'}}\cos\phi_P{\cal F}^{B \rightarrow
a_2}(m^2_{\eta^{'}})/2$  & $1.98 \times 10^{-9}$\\
& $\overline{B^0} \rightarrow \eta^{'} f_2$ &
$a_2f_{\eta^{'}}\cos\phi_P\cos\phi_T{\cal F}^{B \rightarrow
f_2}(m^2_{\eta^{'}})/2$ & $2.13 \times 10^{-9}$ \\
& $\overline{B^0} \rightarrow \eta^{'} f_2^{'}$ & $a_2
f_{\eta^{'}}\cos\phi_P\sin\phi_T{\cal F}^{B \rightarrow
f_2^{'}}(m^2_{\eta^{'}})/2$ & $2.08 \times 10^{-11}$\\
\hline\hline
& Process & Amplitude $\times (V_{ub}V_{cs}^*)$ &\\
\hline\hline
& $B^- \rightarrow D_s^- a_2^0$ & $a_1f_{D_s^-}{\cal F}^{B \rightarrow
a_2}(m^2_{D_s^-})/\sqrt{2}$ & $1.45 \times 10^{-6}$ \\
& $B^- \rightarrow D_s^- f_2$ & $a_1f_{D_s^-}\cos\phi_T{\cal F}^{B
\rightarrow f_2}(m^2_{D_s^-})/\sqrt{2}$ & $1.58 \times 10^{-6}$\\
$\Delta s=1$ & $B^- \rightarrow D_s^-f_2^{'}$ & $a_1f_{D_s^-}\sin\phi_T
{\cal F}^{B \rightarrow f_2^{'}}(m^2_{D_s^-})/\sqrt{2}$ & $1.43 \times
10^{-8}$\\
& $B^- \rightarrow \overline{D^0}K_2^{*-}$ & $a_2f_{D^0}{\cal F}^{B
 \rightarrow K_2^*}(m^2_{D^0})$ & $6.38 \times 10^{-8}$\\ &
 $\overline{B^0} \rightarrow D_s^- a_2^+$ & $a_1f_{D_s^-}{\cal F}^{B
\rightarrow a_2}(m^2_{D_s^-})$ & $2.75 \times 10^{-6}$ \\
& $\overline{B^0} \rightarrow \overline{D^0} \overline{K}_2^{*0}$ &
$a_2f_{D^0}{\cal F}^{B \rightarrow K_2^*}(m^2_{D^0})$ & $5.90 \times
10^{-8}$ \\
\hline
\end{tabular}
\end{center}
\begin{center}
 Table 1. Decay amplitudes and branching ratios for the CKM-suppressed $B
 \rightarrow PT$ channels of the type-I with $\Delta s=0, -1$. These
 amplitudes must be multiplied by $(iG_F/\sqrt{2})\varepsilon^*_{\mu
\nu}p_B^{\mu}p_B^{\nu}$.
\end{center}

\newpage

\begin{center}
\begin{tabular}{|c|c|c|c|}
\hline
 & Process & Amplitude $\times (V_{cb}V_{cd}^*)$ & $Br(B \rightarrow PT)$
\\
\hline\hline
 & $B^- \rightarrow D^- D_2^{*0}$ & $a_1f_{D^-} {\cal F}^{B \rightarrow
D_2^*}(m^2_{D^-})$  & $1.76 \times 10^{-5}$\\
& $B^- \rightarrow \eta_c a_2^-$ & $a_2f_{\eta_c}{\cal F}^{B \rightarrow
a_2}(m_{\eta_c}^2) $ & $1.44 \times 10^{-6}$ \\
$\Delta s=0$ & $\overline{B^0} \rightarrow D^-D_2^{*+}$ &
$a_1f_{D^-}{\cal F}^{B \rightarrow D_2^*}(m^2_{D^-})$ & $1.66 \times
10^{-5}$ \\
& $\overline{B^0} \rightarrow \eta_c a_2^0$ & $-a_2f_{\eta_c}{\cal F}^{B
\rightarrow a_2}(m^2_{\eta_c})/\sqrt{2}$ & $6.80 \times 10^{-7}$\\
& $\overline{B^0} \rightarrow \eta_c f_2$ & $a_2f_{\eta_c}\cos\phi_T{\cal
F}^{B \rightarrow f_2}(m^2_{\eta_c})/\sqrt{2}$ & $7.77 \times 10^{-7}$ \\
& $\overline{B^0} \rightarrow \eta_c f_2^{'}$ &
$a_2f_{\eta_c}\sin\phi_T{\cal F}^{B \rightarrow
f_2^{'}}(m^2_{\eta_c})/\sqrt{2}$ & $5.07 \times 10^{-9}$\\
\hline\hline
& Process & Amplitude $\times (V_{cb}V_{us}^*)$ & \\
\hline\hline
& $B^- \rightarrow K^-D_2^{*0}$ & $a_1f_{K^-}{\cal F}^{B \rightarrow
D_2^*}(m^2_{K^-})$ & $2.40 \times 10^{-5}$ \\
$\Delta s=-1$ & $B^- \rightarrow D^0K_2^{*-}$ & $a_2f_{D^0}{\cal F}^{B
\rightarrow K_2^*}(m^2_{D^0})$ & $4.56 \times 10^{-7}$ \\
& $\overline{B^0} \rightarrow K^-D_2^{*+}$ & $a_1f_{K^-}{\cal F}^{B
\rightarrow D_2^*}(m^2_{K^-})$ & $2.27 \times 10^{-5}$\\
& $\overline{B^0} \rightarrow D^0 \overline{K}_2^{*0}$ & $a_2f_{D^0}{\cal
F}^{B
\rightarrow K_2^{*}}(m^2_{D^0})$ & $4.22 \times 10^{-7}$\\
\hline
\end{tabular}
\end{center}
\begin{center}
Table 2. Decay amplitudes and branching ratios for the CKM-suppressed $B
\rightarrow PT$ decays of the type-II with $\Delta s=0,-1$. The
amplitudes must be multiplied by $(iG_F/\sqrt{2})\varepsilon^*_{\mu
\nu}p_B^{\mu}p_B^{\nu}$. 
\end{center}

\newpage

\begin{center}
\begin{tabular}{|c|c|c|c|}
\hline
& Process & Amplitude $\times (V_{ub}V_{ud}^*)$ & $Br(B \rightarrow VT)$ \\
\hline\hline
& $B^- \rightarrow \rho^- a_2^0$ & $a_1f_{\rho^-}m^2_{\rho^-}{\cal
F}_{\mu \nu}^{B \rightarrow a_2}(m^2_{\rho^-})/\sqrt{2}$ & $8.66 \times
10^{-7}$ \\
& $B^- \rightarrow \rho^- f_2$ &
$a_1f_{\rho^-}m^2_{\rho^-}\cos\phi_T{\cal F}_{\mu \nu}^{B \rightarrow
f_2}(m^2_{\rho^-})/\sqrt{2} $ & $9.22 \times 10^{-7}$\\
& $B^- \rightarrow \rho^- f_2^{'}$ &
$a_1f_{\rho^-}m^2_{\rho^-}\sin\phi_T{\cal F}^{B \rightarrow f_2^{'}}_{\mu
\nu}(m^2_{\rho^-})/\sqrt{2}$ & $9.83 \times 10^{-9}$\\
& $B^- \rightarrow \rho^0 a_2^-$ & $a_2f_{\rho^0}m^2_{\rho^0}{\cal F}^{B
\rightarrow a_2}_{\mu \nu}(m^2_{\rho^0})/\sqrt{2}$ & $4.43 \times
10^{-8}$ \\
& $B^- \rightarrow \omega a_2^-$ &
$a_2f_{\omega}m^2_{\omega}\cos\phi_V{\cal F}^{B \rightarrow a_2}_{\mu
\nu}(m^2_{\omega})/\sqrt{2}$ & $3.59 \times 10^{-8}$\\
& $B ^- \rightarrow \phi a_2^-$ & $a_2f_{\phi}m^2_{\phi}\sin\phi_V{\cal
F}_{\mu \nu}^{B \rightarrow a_2}(m^2_{\phi})/\sqrt{2}$ & $2.31 \times
10^{-10}$ \\
& $\overline{B^0} \rightarrow \rho^- a_2^+$ &
$a_1f_{\rho^-}m^2_{\rho^-}{\cal F}^{B \rightarrow a_2}_{\mu
\nu}(m^2_{\rho^-})$ & $1.64 \times 10^{-6}$ \\
$\Delta s=0$ & $\overline{B^0} \rightarrow \rho^0 a_2^0$ &
$-a_2f_{\rho^0}m^2_{\rho^0}{\cal F}^{B \rightarrow a_2}_{\mu
\nu}(m^2_{\rho^0})/2$ & $2.09 \times 10^{-8}$\\
& $\overline{B^0} \rightarrow \rho^0 f_2$ &
$a_2f_{\rho^0}m^2_{\rho^0}\cos\phi_T{\cal F}^{B \rightarrow f_2}_{\mu
\nu}(m^2_{\rho^0})/2$ & $2.23 \times 10^{-8}$ \\
& $\overline{B^0} \rightarrow \rho^0 f_2^{'}$ &
$a_2f_{\rho^0}m^2_{\rho^0}\sin\phi_T{\cal F}^{B \rightarrow f_2^{'}}_{\mu
\nu}(m^2_{\rho^0})/2$ & $2.38 \times 10^{-10}$\\
& $\overline{B^0} \rightarrow \omega a_2^0$ &
$-a_2f_{\omega}m^2_{\omega}\cos\phi_V{\cal F}^{B \rightarrow a_2}_{\mu
\nu}(m^2_{\omega})/2$ & $1.70 \times 10^{-8}$\\
& $\overline{B^0} \rightarrow \omega f_2$ &
$a_2f_{\omega}m^2_{\omega}\cos\phi_V\cos\phi_T{\cal F}^{B \rightarrow
f_2}_{\mu \nu}(m^2_{\omega})/2$ & $2.42 \times 10^{-8}$\\
& $\overline{B^0} \rightarrow \omega f_2^{'}$ &
$a_2f_{\omega}m^2_{\omega}\cos\phi_V\sin\phi_T{\cal F}^{B \rightarrow
f_2^{'}}_{\mu \nu}(m^2_{\omega})/2$ & $1.93 \times 10^{-10}$\\
& $\overline{B^0} \rightarrow \phi a_2^0$ &
$-a_2f_{\phi}m^2_{\phi}\sin\phi_V{\cal F}^{B \rightarrow a_2}_{\mu
\nu}(m^2_{\phi})/2$ & $1.09 \times 10^{-10}$\\
& $\overline{B^0} \rightarrow \phi f_2$ &
$a_2f_{\phi}m^2_{\phi}\sin\phi_V\cos\phi_T{\cal F}^{B \rightarrow
f_2}_{\mu \nu}(m^2_{\phi})/2$ & $1.16 \times 10^{-10}$ \\ 
& $\overline{B^0} \rightarrow \phi f_2^{'}$ &
$a_2f_{\phi}m^2_{\phi}\sin\phi_V\sin\phi_T{\cal F}^{B \rightarrow
f_2^{'}}_{\mu \nu}(m^2_{\phi})/2$ & $1.29 \times 10^{-12}$\\
\hline\hline
& Process & Amplitude $\times (V_{ub}V_{cs}^*)$ & \\
\hline\hline
& $B^- \rightarrow D_s^{*-}a_2^0$ & $a_1f_{D_s^{*-}}m^2_{D_s^{*-}}{\cal
F}^{B \rightarrow a_2}_{\mu \nu}(m^2_{D_s^{*-}})/\sqrt{2}$ & $2.13 \times
10^{-6}$ \\
& $B^- \rightarrow D_s^{*-} f_2$ &
$a_1f_{D_s^{*-}}m^2_{D_s^{*-}}\cos\phi_T{\cal F}^{B \rightarrow f_2}_{\mu
\nu}(m^2_{D_s^{*-}})/\sqrt{2}$ & $2.17 \times 10^{-6}$\\
$\Delta s=-1$ & $B^- \rightarrow D_s^{*-}f_2^{'}$ &
$a_1f_{D_s^{*-}}m^2_{D_s^{*-}}\sin\phi_T{\cal F}^{B \rightarrow
f_2^{'}}_{\mu \nu}(m^2_{D_s^{*-}})/\sqrt{2}$ & $2.98 \times 10^{-8}$\\
& $B^- \rightarrow \overline{D^{*0}} K_2^{*-}$ &
$a_2f_{D^{*0}}m^2_{D^{*0}}{\cal F}^{B \rightarrow K_2^{*-}}_{\mu
\nu}(m^2_{D^{*0}})$ & $3.14 \times 10^{-7}$\\
& $\overline{B^0} \rightarrow D_s^{*-} a_2^+$ &
$a_1f_{D_s^{*-}}m^2_{D_s^{*-}}{\cal F}^{B \rightarrow a_2}_{\mu
\nu}(m^2_{D_s^{*-}})$ & $4.03 \times 10^{-6}$\\
& $\overline{B^0} \rightarrow \overline{D^{*0}} \overline{K}_2^{*0}$ &
$a_2f_{D^{*0}}m^2_{D^{*0}}{\cal F}^{B \rightarrow K_2^*}_{\mu
\nu}(m^2_{D^{*0}})$ & $2.95 \times 10^{-7}$ \\
\hline
\end{tabular}
\end{center}
\begin{center}
Table 3. Decay amplitudes and branching ratios for the CKM-suppressed $B
\rightarrow VT$ modes of the type-I with $\Delta s=0,-1$. The amplitudes
must be multiplied by $(G_F/\sqrt{2})\varepsilon^{*\mu \nu}$.
\end{center}

\newpage

\begin{center}
\begin{tabular}{|c|c|c|c|}
\hline
& Process & Amplitude $\times (V_{cb}V_{cd}^*)$ & $Br(B \rightarrow VT)$\\
\hline\hline
& $B^- \rightarrow D^{*-}D_2^{*0}$ & $a_1f_{D^{*-}}m^2_{D^{*-}}{\cal
F}^{B \rightarrow D_2^*}_{\mu \nu}(m^2_{D^{*-}})$ & $6.66\times 10^{-5}$ \\
& $B^- \rightarrow J/\psi a_2^-$ & $a_2f_{J/\psi}m^2_{J/\psi}{\cal F}^{B
\rightarrow a_2}_{\mu \nu}(m^2_{J/\psi})$ & $5.23 \times 10^{-6}$\\
$\Delta s=0$ & $\overline{B^0} \rightarrow D^{*-} D_2^{*+}$ &
$a_1f_{D^{*-}}m^2_{D^{*-}}{\cal F}^{B \rightarrow D_2^*}_{\mu
\nu}(m^2_{D^{*-}})$ & $6.29 \times 10^{-5}$ \\
& $\overline{B^0} \rightarrow J/\psi a_2^0$ &
$-a_2f_{J/\psi}m^2_{J/\psi}{\cal F}^{B \rightarrow a_2}_{\mu
\nu}(m^2_{J/\psi})/\sqrt{2}$ & $2.47 \times 10^{-6}$ \\
& $\overline{B^0} \rightarrow J/\psi f_2$ &
$a_2f_{J/\psi}m^2_{J^/\psi}\cos\phi_T{\cal F}^{B \rightarrow f_2}_{\mu
\nu}(m^2_{J/\psi})/\sqrt{2}$ & $2.56 \times 10^{-6}$ \\
& $\overline{B^0} \rightarrow J/\psi f_2^{'}$ &
$a_2f_{J/\psi}m^2_{J/\psi}\sin\phi_T{\cal F}^{B \rightarrow f_2^{'}}_{\mu
\nu}(m^2_{J/\psi})/\sqrt{2}$ & $2.92 \times 10^{-8}$\\
 \hline\hline
& Process & Amplitude $\times (V_{cb}V_{us}^*)$ & \\
\hline\hline
& $B^- \rightarrow K^{*-} D_2^{*0}$ & $a_1f_{K^{*-}}m^2_{K^{*-}}{\cal
F}^{B \rightarrow D_2^*}_{\mu \nu}(m^2_{K^{*-}})$ & $5.77 \times
10^{-5}$\\
$\Delta s=-1$ & $B^- \rightarrow D^{*0}K_2^{*-}$ &
$a_2f_{D^{*0}}m^2_{D^{*0}}{\cal F}^{B \rightarrow K_2^*}_{\mu
\nu}(m^2_{D^{*0}})$  & $2.24 \times 10^{-6}$\\
& $\overline{B^0} \rightarrow K^{*-}D_2^{*+}$ &
$a_1f_{K^{*-}}m^2_{K^{*-}}{\cal F}^{B \rightarrow D_2^*}_{\mu
\nu}(m^2_{K^{*-}})$  & $5.46 \times 10^{-5}$\\
& $\overline{B^0} \rightarrow D^{*0} \overline{K}_2^{*0}$ &
$a_2f_{D^{*0}}m^2_{D^{*0}}{\cal F}^{B \rightarrow K_2^*}_{\mu
\nu}(m^2_{D^{*0}})$ & $2.11 \times 10^{-6}$\\
\hline
\end{tabular}
\end{center}
\begin{center}
Table 4. Decay amplitudes and branching ratios for the CKM-suppressed $B
\rightarrow VT$ decays of the type-II with $\Delta s=0,-1$. The
amplitudes must be multiplied by $(G_F/\sqrt{2})\varepsilon^{*\mu \nu}$. 
\end{center}

\newpage

\begin{center}
\begin{tabular}{|c|c|c|c|}
\hline
& Process & Amplitude $\times (V_{cd}V_{ud}^*)$ & $Br(D \rightarrow PT)$\\
\hline\hline
& $D^0 \rightarrow \pi^+ a_2^-$ & $a_1f_{\pi^+}{\cal F}^{D \rightarrow
a_2}(m^2_{\pi^+})$ & $4.21 \times 10^{-6}$ \\
& $D^0 \rightarrow \pi^0 a_2^0$ & $-a_2f_{\pi^0}{\cal F}^{D \rightarrow
a_2}(m^2_{\pi^0})/2$ & $1.72 \times 10^{-7}$\\
& $D^0 \rightarrow \pi^0 f_2$ & $-a_2f_{\pi^0}\cos\phi_T{\cal F}^{D
\rightarrow f_2}(m^2{\pi^0})/2$ & $2.47 \times 10^{-7}$ \\
$\Delta s=0$ & $D^0 \rightarrow \pi^0 f_2^{'}$ & $-a_2f_{\pi^0}\sin\phi_T
{\cal F}^{D \rightarrow f_2^{'}}(m^2_{\pi^0})/2$ & $2.18 \times 10^{-10}$\\
& $D^+ \rightarrow \pi^+ a_2^0$ & $-a_1f_{\pi^+}{\cal F}^{D \rightarrow
a_2}(m^2_{\pi^+})/\sqrt{2}$ & $5.55 \times 10^{-6}$\\
& $D^+ \rightarrow \pi^+ f_2$ & $a_1f_{\pi^+}\cos\phi_T{\cal F}^{D
\rightarrow f_2}(m^2_{\pi^+})/\sqrt{2}$ & $7.97 \times 10^{-6}$\\
& $D^+ \rightarrow \pi^+ f_2^{'}$ & $a_1f_{\pi^+}\sin\phi_T{\cal F}^{D
\rightarrow f_2^{'}}(m^2_{\pi^+})/\sqrt{2}$  & $7.18\times 10^{-9}$\\
& $D^+ \rightarrow \pi^0 a_2^+$ & $-a_2f_{\pi^0}{\cal F}^{D \rightarrow
a_2}(m^2_{\pi^0})/\sqrt{2}$ & $9.05 \times 10^{-7}$\\
\hline

\end{tabular}
\end{center}
\begin{center}
Table 5. Decay amplitudes and branching ratios for the CKM-suppressed $D
\rightarrow PT$ channels of the type-I with $\Delta s=0$. The amplitudes
must be multiplied by $(iG_F/\sqrt{2})\varepsilon^*_{\mu
\nu}p_D^{\mu}p_D^{\nu}$. 
\end{center}

\end{document}